\documentclass[10pt]{article}
\usepackage[footnotesize]{caption}

\usepackage{float}
\usepackage[symbol]{footmisc}
\usepackage[superscript,sort]{cite}
\usepackage{array}
\usepackage[pdftex,colorlinks=true,linkcolor=black,citecolor=black,urlcolor=black,bookmarks=true,breaklinks=true]{hyperref}
\usepackage[pdftex]{graphicx}
\usepackage[usenames,pdftex]{color}
\definecolor{Red}{rgb}{1.0,0.0,0.0}
\usepackage{amsmath,amssymb}
\usepackage{amsthm,amscd,amsxtra,amsfonts,amsmath,amssymb,multirow}
\usepackage{wrapfig}
\usepackage[tiny,compact]{titlesec}
\usepackage{helvet}

\usepackage{xcolor,colortbl}
\usepackage{subcaption}
\usepackage{algorithm}
\usepackage[noend]{algpseudocode}
\usepackage{tikz}
\usepackage[T1]{fontenc}

\titlespacing*{\section}{0pt}{*0}{*0}
\titlespacing*{\subsection}{0pt}{*0}{*0}
\titlespacing*{\subsubsection}{0pt}{*0}{*0}
\titlespacing{\paragraph}{0pt}{*0}{*1}

\definecolor{MyPurple}{rgb}{1,0,1}

\newcommand{\beq}[1]{\begin{equation} \label{#1}}
\newcommand{\eeq}{\end{equation}}
\newcommand{\barray}{\begin{array}{ll}}
\newcommand{\earray}{\end{array}}

\begin{document}
\pagenumbering{roman}

\clearpage \pagebreak \setcounter{page}{1}
\renewcommand{\thepage}{{\arabic{page}}}

\title{Comparison of  multi-task convolutional neural network (MT-CNN) and  a few other methods for toxicity prediction}

\author{Kedi Wu$^{1}$ and  Guo-Wei Wei$^{1,2,3}$
\footnote{
Address correspondences  to Guo-Wei Wei. E-mail: wei@math.msu.edu}  \\
$^1$Department of Mathematics \\
Michigan State University, MI 48824, USA\\
$^2$Department of Electrical and Computer Engineering \\
Michigan State University, MI 48824, USA \\
$^3$Department of Biochemistry and Molecular Biology\\
Michigan State University, MI 48824, USA
}

\date{\today}
\maketitle

\begin{abstract}
Toxicity analysis and prediction are of paramount importance to  human health and environmental protection. Existing computational methods are built from a wide variety of descriptors and regressors, which makes their performance analysis difficult. For example, deep neural network (DNN),  a successful approach in many occasions, acts like a black box and offers little conceptual elegance or physical understanding. The present work constructs a common set of microscopic  descriptors based on established physical models for charges, surface areas and free energies to assess the performance of multi-task convolutional  neural network (MT-CNN) architectures and a few other approaches, including random forest (RF) and gradient boosting decision tree (GBDT), on an equal footing.  Comparison is also given to  convolutional neural network (CNN) and non-convolutional deep neural network (DNN) algorithms. Four benchmark toxicity data sets (i.e., endpoints) are used  to evaluate various approaches. Extensive numerical studies indicate that the present MT-CNN architecture is able to outperform the state-of-the-art methods. 

\end{abstract}

Key words: toxicity endpoints, multitask learning, deep neural network.
\maketitle

\newpage
{\setcounter{tocdepth}{4} \tableofcontents}
\newpage

\section{Introduction}\label{sec:Intro}

Toxicity is a measure of the degree to which a chemical can adversely affect an organism. These adverse effects, which are called toxicity endpoints, can be quantitatively or qualitatively measured by their effects on given targets.   Most toxicity tests aim to protest human from harmful effects caused by chemical substances and are traditionally conducted in {\sl in vivo} or {\sl in vitro} manner. Nevertheless, such experiments are usually very time consuming and cost intensive, and even give rise to ethical concerns when it comes to animal tests. Therefore, computer-aided methods, or {\sl in silico} methods, have been developed to improve prediction efficiency without sacrificing too much of accuracy. Quantitative structure activity relationship (QSAR) approach is one of the most popular and commonly used approaches. The basic QASR assumption is that similar molecules have similar activities. Therefore by studying the relationship between chemical structures and biological activities, it is possible to predict the activities of new molecules without actually conducting lab experiments. Thanks to the development of high throughput screening (HTS) techniques, researchers are now able to quickly and routinely handle thousands of chemicals, which in turn provides high volume data for drug discovery and for training and validating QSAR models. 

A major concern for developing QSAR models is to choose appropriate chemical features (descriptors). On one hand,  QSAR models are easy to interpret if molecular features are well studied and are of chemical or physical meanings. Using excessive or insufficient descriptors, however, can make interpretation of QSAR models very difficult. The general feature selection process may cause problems in situations where statistically important features may not be biologically interpretable. On the other hand, the development of accurate QSAR models demands sufficiently large number of training chemicals and reasonable external validation sets. A QSAR model is likely to overfit data and fails to give satisfactory predictions if limited amount of experimental data is available. In addition, QSAR models that involve a large set of descriptors  may frequently encounter chemicals of particular interest that are out of QSAR's applicable domain because one or more descriptors cannot be constructed. In such cases, new models have to be developed to make an accurate prediction. 

There are several types of algorithms to generate QSAR models: linear models based on linear regression and linear discriminant analysis \cite{deeb:2012};  nonlinear models including nearest neighbor \cite{kauffman:2001, ajmani:2006}, support vector machine  \cite{deeb:2012,si:2007, du:2008} and random forest \cite{svetnik:2003}. These methods have advantages and disadvantages \cite{liupx:2009} due to their statistics natures. For instance, linear models overlook the relatedness between different features, while nearest neighbor method largely depends on the choice of descriptors. To overcome these difficulties, more refined and advanced machine learning methods have been introduced. Multi-task learning (MTL)  \cite{caruana:1998} was proposed partially to deal with  aforementioned data sparsity problem. The idea of MTL is to learn the so-called `inductive bias' from related tasks to improve accuracy using the same representation. In other words, MTL aims at learning a shared and generalized feature representation from multiple tasks. It is therefore promising to learn shared information across different tasks and give better predictions. Indeed, MTL strategies have brought new insights to bioinformatics since compounds from related assays may share features at various feature levels, which is extremely helpful if data set is small. Successful applications include splice-site and MHC-I binding prediction \cite{widmer:2012} in sequence biology, gene expression analysis, and system biology \cite{xuq:2011}. 

Recently, deep learning (DL) \cite{lecun:2015,schmidhuber:2015}, particularly convolutional neural network (CNN), has emerged as powerful paradigm to render a wide range of state-of-art results in signal and information processing fields, such as speech recognition \cite{dahl:2012, deng:2013} and natural language processing  \cite{socher:2012, sutskever:2014}. Deep learning architecture is essentially based on artificial neural networks. The major difference between deep neural network (DNN)  models and non-DNN models is that DNN models consist of a large number of layers and neurons, making it possible to construct abstract features in each layer and pass higher level information to next layer. For instance, in face recognition, DNs detect blobs and edges from raw  unprocessed images, and further reveal objects such as eyes and noses. As a result, DNNs can easily be applied to QSAR framework. Given a molecular feature representation, DNNs are able to capture abstract chemical features from raw  unprocessed  data. Dahl {\it et al.}  \cite{dahl:2014} used artificial neural network to predict activities of compounds for multiple assays at the same time. Ma {\it et al.}  \cite{majinshui:2015} from Merck discussed the optimization of parameters for single and multitask neural nets and improved prediction accuracy over random forest. Another related work was done by Mayr {\it et al.}  \cite{mayr:2016} in Tox21 Data challenge, where they trained a multitask network on a dataset of 1,280 biological targets with 2 million associated data points.  Ramsundar {\it et al.}  \cite{ramsundar:2015} further investigated the relationship between the amount of data and the number of tasks, and between the predictive power and transferability of multitask neural networks.

However, there is typically no physical interpretation  of  DNN results. Indeed, DNN approaches offer little conceptual understanding. In contrast, physical based methods provide conceptual elegance in terms of  first principles, i.e., fundamental laws of physics. Additionally, DNNs behave    mysteriously to the designer and the user alike. Once created, not even the designer can explain what exactly the neural network has learned at each layer and how the property of interest is predicted finally. Consequently, the design of DNNs is often based on trial and error, and mostly limited by accessible computer power. Moreover, for a given method, features or descriptors have a major impact on prediction accuracy.  Most of early comparisons of DNN and non-DNN methods are not only based on different methods, but also based on different features. It is meaningful to compare the performance of different methods with a common set of features.    
   
Objective of this work is to design a common set of {\it  microscopic  features} based on established physical models so that DL models,  including multi-task CNN (MT-CNN), MT-DNN, single-task CNN (ST-CNN) and ST-DNN and two non-DL approaches can be compared on an equal footing. These physical features have led to excellent protein-ligand binding affinity predictions in  our earlier work \cite{BaoWang:2016FFTB}.
To extract physical features from raw data, we first set up three-dimensional (3D) molecular models for all compounds. Appropriate force field that was originally derived from quantum mechanics is utilized to obtain atomic charges. Atomic surface areas are calculation based an appropriate geometric model. Finally,  atomic electrostatic solvation free energies are computed via the Poisson model. These physical features and their statistical quantities allow us to construct a number of efficient DL and non-DL models for toxicity predictions.

The paper is structured as following. Section \ref{sec:methods} is devoted to methods and algorithms.   Feature generation and element-wise feature construction are described in detail. A brief overview of classic ensemble methods and a detailed description of deep learning  architectures are given.    Single-task and multi-task deep neural network architectures are discussed.   In Section \ref{sec:data},  we give a description on data sets used to develop QSAR models, followed by model parametrization and evaluation criteria. Section  \ref{sec:results} presents our prediction results for four benchmark  toxicity endpoints, followed by Section \ref{sec:discussion} which puts an emphasis on model optimization. Finally, this paper is wrapped up with some concluding remarks in Section  \ref{sec:conclusion}.

\section{Methods and algorithms} \label{sec:methods}

In this section, we first provide a detail description of feature generation. Then, an overview of ensemble methods (random forest and gradient boosting decision tree),  deep neural networks, single-task learning and multi-task learning is given. Emphasis will be put on advantages of multi-task deep convolutional neural network for quantitative toxicity endpoint predictions and how to select appropriate parameters for network architectures. A detailed description of our architecture and training procedure is also provided.

\subsection{Physical feature generation} \label{sec:feature}

In this work, we are interested in constructing a set of high level microscopic features based physical models to describe  molecular toxicity. This set of features should be  convenient for being used in different machine learning approaches, including deep learning and non deep learning, and single-task and multi-task ones. One might argue that it is unfair to feed deep CNN with high level physical features. However, it is worthy to note that unlike image classification problems, for chemical and biological data sets, the raw image based data input does not work for CNN because the chemical information cannot be directly represented by raw images. Therefore, certain level of data preprocessing is always required in the application of CNN to chemical and biological data.  

To make our feature generation feasible and robust to all compounds, we consider three types of basic physical information, i.e.,, atomic charges computed from quantum mechanics or molecular force fields, atomic surface areas calculated for solvent excluded surface definition, and atomic electrostatic solvation free energies estimated from the Poisson model.  To obtain this information, we first construct optimized 3D structure of for each molecule. Then the aforementioned atomic properties are computed. Our feature  generation process can be divided into several steps:
\begin{enumerate}
\item {\bf Structure} Optimized 3D structures were prepared by LigPrep in \href{https://www.schrodinger.com/}{ Schr\"{ o}dinger suites (2014-2)}  from the original 2D structures, using options: \{-i 0 -nt -s 10 -bff 10\}. 

\item {\bf Charge} Optimized 3D structures were then fed in antechamber  \cite{wangjm:2006}, using parametrization: AM1-BCC charge, Amber mbondi2 radii and general Amber force field (GAFF) \cite{wangjm:2004}. This step leads to pqr files with corresponding charge assignments.

\item  {\bf Surface} \href{http://weilab.math.msu.edu/ESES/}{ESES online server}  \cite{ESES:2017}  was used to compute atomic surface area of each molecule, using pqr files from the previous step. This step also results in molecular solvent excluded surface information.  

\item {\bf Energy} \href{http://weilab.math.msu.edu/MIBPB/}{MIBPB online server}  \cite{DuanChen:2011a} was used to calculate the atomic electrostatic solvation free  energy  of each molecule, using surface and pqr files from previous steps. 
\end{enumerate}  
Physical features were obtained according according the above procedure. Specifically, our features comes from step 2, step 3 and step 4.  Unlike other traditional QSAR approaches which incorporate 2D macroscopic descriptors such as ECFP4 fingerprints or the  number of certain atom types, we put emphasis on the physical understanding of toxicity via atomic features generated from physical models. 
However, atomic features are associated with atomic positions, which  cannot be directly compared among different molecules. Additionally molecules typically have different number of atoms. Therefore, a direct comparison of atomic level information is not feasible for most machine learning algorithms.  We construct element features, in which atomic properties of the same element type are summed together. Consequently, we have element charge, surface area and electrostatic solvation free energies. Additionally, for element feature, we further consider their statistics, i.e., maximum, minimum, mean and variance, which gives rise  to another four features. Moreover,  strengthens of atomic charge and  atomic electrostatic solvation free energies, obtained from absolute values, are important physical quantities. Therefore,  taking absolute value of  atomic charges and electrostatic solvation free energies before the summations, we generate element charge strength feature and  electrostatic solvation strength feature for each type of element.   
For each element strength feature, its statistical quantities, i.e., maximum, minimum, mean and variance, lead to another four features.
Finally, for the sake of symmetry among original three types of physical quantities, we create five strength features for element  surface ares as well. This symmetry consideration is important for our convolutional algorithm and its redundancy is essentially harmless to all machine learning methods. Consequently, we have a total of 30 element level physical features for each type of element.   In order for our physical models to be applicable to a wide variety of chemicals, we only consider 10 different commonly occurring element types, i.e., {\rm H, C, N, O, F, P, S, Cl, Br,} and {\rm I}, in our element selection.

In this work, we also construct coordinate-free molecular features from atomic charges, surface areas and electrostatic solvation free energies in a similar manner. This consideration gives rise to  30 molecular features.  We organize this element specific and molecular specific information into 11 channels, which are analogous to RGB channels in the RBG color image representation. Consequently, the final input feature matrix for a given molecule has the dimension of $11\times30$.  In order to achieve numerical stability and faster convergence (a larger learning rate) in CNN, the input matrix is scaled along the feature dimension.  Namely, for each feature, it is normalized to zero mean and unit variance. The resulting input matrix is readily suitable for CNN. 

Alternatively, we organize all of the above information into a 1D feature vector with 330 components, which is readily suitable  for ensemble methods, such as random forest and gradient boosting decision tree, and for non-convolutional DNN.

\subsection{Ensemble methods} 

To explore strengths and weaknesses of different machine learning methods, we consider two popular ensamble methods, namely,  random forest (RF) and gradient boosting decision tree (GBDT). These  approaches have been widely used in solving QSAR prediction problems, as well as solvation and protein-ligand binding free energy predictions \cite{BaoWang:2016FFTS,BaoWang:2016FFTB,ZXCang:2017b}. They naturally  handle correlation between descriptors, and usually do not require a sophisticated feature selection procedure. Most importantly, both RF and GBDT are essentially  insensitive to parameters. Therefore,  we choose these two machine learning methods as baselines in our comparison. 

We have implemented these regressors using the scikit-learn package (version 0.13.1)  \cite{scikit-learn}. The number of estimators is set to $2000$ and the learning rate is optimized for GBDT method. For each set, 50 runs (with different random states) were done and the average result is reported in this work.  The 1D feature vector with 330 components discussed in Section \ref{sec:feature} is used as input data for RF and GBDT. 

\subsection{Single-task deep learning algorithms  } \label{sec:CNN_arch}
A neural network acts as a transformation  that maps an input feature vector to an output vector. It essentially models the way a biological brain solves problems with numerous neuron units connected by axons. A typical shallow neural network consists of a few layers with neurons and uses back propogation to update weights on each layer. However, it is not able to construct hierarchical features and thus falls short in revealing more abstract properties, which makes it difficult to model complex non linear relationships.  

A single-task deep learning algorithm, compared to shallow networks, has a wider and deeper architecture --  it consists of more layers and more neurons in each layer and reveals the facets of input features at different levels. Single-task deep learning algorithm is defined for each individual prediction task and only learns data from the specific task.

\begin{figure}[!ht]
\small
\centering
\includegraphics[scale=0.8]{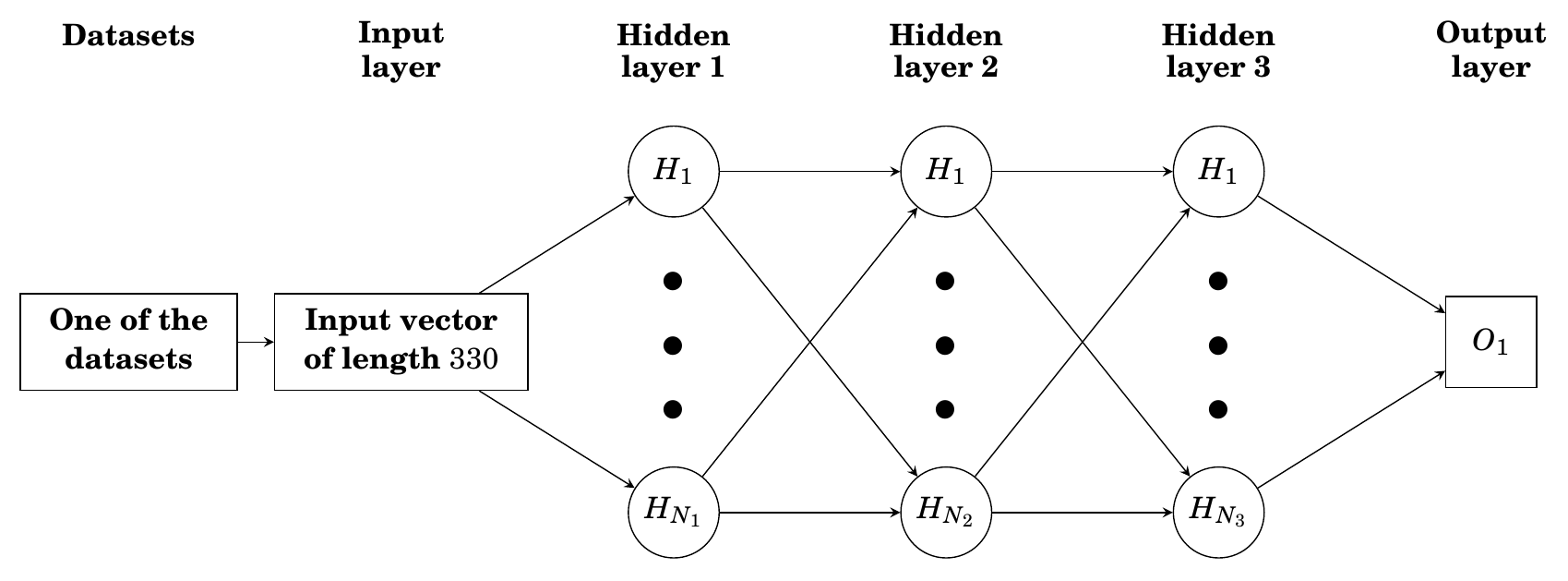}
\caption{An illustration of the ST-DNN architecture.}
\label{fig:st-DNN_arch}
\end{figure}
 In this work, we construct two types of single-task  (ST) deep learning algorithms, i.e., ST-CNN and ST-DNN. The 1D feature vector with 330 components discussed in Section \ref{sec:feature} is used as input data for ST-DNN, while the feature matrix of dimension $11\times30$ is used for ST-CNN. A representation of such ST-DNN can be found in Fig. \ref{fig:st-DNN_arch}, where $N_i$ $(i=1,2,3)$ represents the number of neurons on the $i$-th hidden layer. 

\subsection{Multi-task learning} \label{sec:MTL}
Multi-task  learning  is a machine learning technique which has shown success in recent Merck and Tox21 prediction challenges. The main advantage of MT learning is to learn multiple tasks simultaneously and exploit commonalities as well as differences across different tasks.  Another advantage of MT learning is that a small data set with incomplete statistical distribution to establish an accurate predictive model can often be significantly benefited from  relatively large data sets with more complete statistical distributions.

 Suppose we have a total of $T$ tasks and the training data for the $t$-th task are denoted as $(\mathbf{x}_i^t, y_i^t)_{i=1}^{N_t}$, where $t={1,..,T}$, $i={1,...,N_t}$, $N_t$ is the number of samples of the $t$-th tasks, with $\mathbf{x}_i^t$ and $\mathbf{y}_i^t$ being the feature vector and target value, respectively. The goal of the MTL is to simultaneously minimize
\begin{equation}
\mathrm{argmin} \sum_{i=1}^{N_t} L(y_i^t, f(\mathbf{x}_i^t; \{\mathbf{W}^t,\mathbf{b}^t\} )) 
\end{equation}
where $f$ is a function of $\mathbf{x}^t$ parametrized by a weight vector $\mathbf{W}^t$ and bias term $\mathbf{b}^t$, and $L$ is the loss function. A typical cost function for regression is the mean squared error, thus the loss of the $t$-th task can be defined as:
\begin{equation}\label{loss}
\mathrm{Loss \hspace{.03in} of \hspace{.03in}}  \mathrm{Task\hspace{.03in} }t = \frac{1}{2}\sum_{i=1}^{N_t} L(\mathbf{x}_i^t, y_i^t) = \frac{1}{2}\sum_{i=1}^{N_t} (y_i^t - f(\mathbf{x}_i^t; \{\mathbf{W}^t,\mathbf{b}^t\} )^2.
\end{equation}
To avoid overfitting problem, we customize the above loss function (\ref{loss}) by adding a regularization term on weight vectors, giving us an improved loss function for $t$-th task:
 \begin{equation}\label{loss2}
\mathrm{Loss \hspace{.03in} of \hspace{.03in}}  \mathrm{Task\hspace{.03in} }t = \frac{1}{2}\sum_{i=1}^{N_t} (y_i^t - f(\mathbf{x}_i^t; \{\mathbf{W}^t,\mathbf{b}^t\} )^2 + \beta||\mathbf{W}^t||_2^2.
\end{equation}
where $||\cdot||$ denotes the $L_2$ norm and $\beta$ represents a penalty constant.

In the context of toxicity prediction, MT learning is to learn different toxicity endpoints jointly and potentially improve the overall performance of multiple regression tasks. More concretely, it is reasonable to assume that different small molecules with different measured toxicity endpoints comprise distinct physical or chemical features, while descriptors, such as the occurrence of certain chemical structure, can result in similar toxicity property.

\subsection{Multi-task convolutional neural network (MT-CNN) architecture} 

To take advantage of MT learning, we propose two MT deep learning architectures --- one makes use of convolutional filters (i.e., MT-CNN) while the other does not (i.e., MT-DNN). In both approaches,  the first few shared layers are designed to capture  a shared representation of different toxicities and the last layer takes care of differences. The proposed neural networks are essentially  nonlinear regressors that perform repeated linear and nonlinear transformations on the input data. 

The main idea of our convolutional architecture is to construct different channels (analogous to RGB channels in image representation) by chemical elements and then perform convolutions on training data along the physical feature dimension constituted with charge, surface area and electrostatic solavtion free energy. The feature extraction stage for different element channels contains one 1D convolutional layer, one max pooling layer, and three more fully connected layers. The convolutional layer contains $60$ filters with size $3$ and stride $1$, whose weights are not spatially shared. It means that a different set of filters is applied at every location of the input vector. The rectify function $\varphi (x)$, which is defined as $\varphi (x) = \max(0,x)$, is chosen as the activation function for the convolutional layer. Then we perform 1D non-overlapping max-pooling of size $2$ on top of the previous layer. Three more fully connected dense layers of various widths are further added to the pooling layer. Note that all above layers are shared across different tasks. Finally in order to capture the differences among different tasks, we attach individual predictor to previous layer for each task.  For regression tasks, a linear function ($f(x)=x$) is sufficient as target values are continuous.  Specifically, our MT-CNN architecture can be represented as:
\begin{itemize}
 \item Input layer [$ 11 \times 30$] holds 30 features for  the molecular channel and 10 element channels.
 \item Convolutional layer computes the output of neurons that are locally connected to the input feature vector. This operation results in a 3D tensor in the form of [$ 11 \times 28 \times 60$], where $28 = 30 -3 +1$, with 3 being the window size of each filter and 1 being the step size when one moves the filter along the feature dimension. Here ``60'' is the number of filters, which can be optimized. 
 \item Rectify layer applies an element-wise activation of $\max(0,x)$, while keeps the volume unchanged.
 \item Maxpooling layer performs a downsampling operation along the feature dimension, resulting in the volume of [$11 \times 14 \times 60$]. 
 \item Fully connected layers output the prediction for each regression task. In each layer, every neuron is connected to all the neurons in the previous layer.
\end{itemize}

\begin{figure}[!ht]
\small
\centering
\includegraphics[scale=0.7]{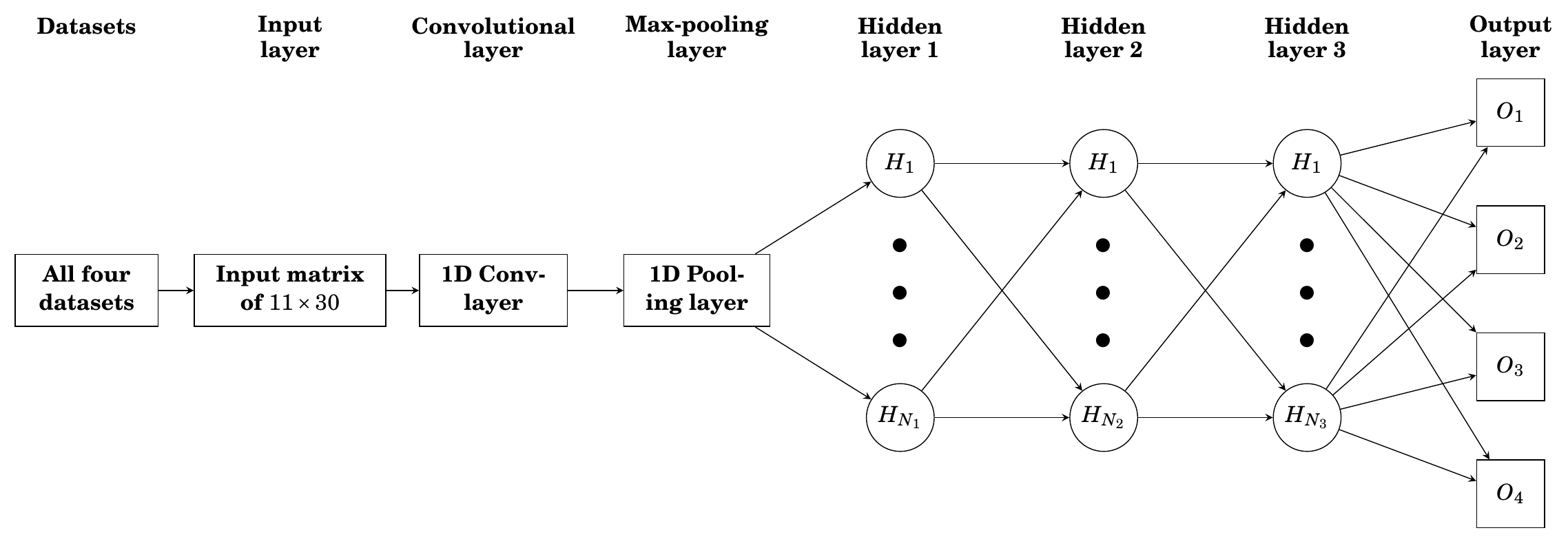}
\caption{An illustration of the convolutional MT-CNN architecture.}
\label{fig:cnn_arch}
\end{figure}

Fig. \ref{fig:cnn_arch} shows the detailed architecture for our convolutional MT-CNN. Here $N_i$ $(i=1,2,3)$ represents the number of neurons on the $i$-th hidden layer. The difference between the convolutional and non-convolutional architectures is that the non-convolutional version has the 1D feature vector with 330 components discussed in Section \ref{sec:feature} as the input data, and it does not have the convolutional layer or max-pooling layer.

\subsection{Model training} 

The training procedure is based on backward propagation (BP) algorithm using mini-batched stochastic descent (SGD) with momentum. Specifically, let $L$ be the objective loss function, $\Theta $ be the parameter to be updated, $v$ be the velocity function, and $<\frac{\partial L}{\partial \Theta}>$ be the average objective function gradient over the current mini-batch, updated rules are as follows:
\begin{align}
   &  v_i = mv_{i-1} - l<\frac{\partial L}{\partial \Theta}> \\
   &  \Theta_i = \Theta_{i-1} + v_i
\end{align}
where $m$ is the momentum and $l$ is the learning rate. 

Glorot weights sampled from uniform distribution are used to initialize weights and bias terms in ({\ref{loss2}}). To reduce prediction bias of the present model, molecules in each training set are randomly shuffled and then divided into mini-batches of size $200$, which are then used to update parameters. When all mini-batches are traversed, an training ``epoch" is done.

\section{Data, parametrization and evaluation } \label{sec:data}

This section is devoted to a detailed description of data sets, model parametrization and result evaluation criteria.  
\subsection{Data sets} \label{sec:data_construction}
Our models were trained and tested on four different datasets -- 96 hour fathead minnow LC$_{50}$ data set (LC$_{50}$ set), 48 hour Daphnia magna LC$_{50}$ data set (LC$_{50}$-DM set), 40 hour Tetrahymena pyriformis IGC$_{50}$ data set (IGC$_{50}$ set), and oral rat LD$_{50}$ data set (LD$_{50}$ set). LC$_{50}$ set aims at predicting the concentration of test chemicals in water in mg/L that causes 50\% of fathead minnow to die after 96 hours. Similarly, LC$_{50}$-DM set seeks to predict the concentration of test chemicals in water in mg/L that causes 50\% Daphnia maga to die after 48 hours. Both sets were originally downloadable from the ECOTOX aquatic toxicity database via web site \url{http://cfpub.epa.gov/ecotox/} and were preprocessed using filter criterion including media type, test location, etc  \cite{test_guide}.  The third set, IGC$_{50}$ set, measures the 50\% growth inhibitory concentration of Tetrahymena pyriformis organism after 40 hours. It was obtained from Schultz and coworkers \cite{akers:1999, zhuhao:2008}. The endpoint LD$_{50}$ represents the amount of chemicals that can kill half of rates when orally ingested. The LD$_{50}$ was constructed from ChemIDplus databse (\url{http://chem.sis.nlm.nih.gov/chemidplus/chemidheavy.jsp}) and then filtered according to several criteria \cite{test_guide}. 

 The final sets used in this work are identical to those that were preprocessed and used to develop the  (\href{https://www.epa.gov/chemical-research/toxicity-estimation-software-tool-test}{Toxicity Estimation Software Tool} (TEST) \cite{test_guide}. TEST was developed to estimate chemical toxicity using various QSAR methodologies and is very convenient to use as it does not require any external programs. It follows the general QSAR workflow --- it first calculates 797 2D molecular descriptors and then predicts the toxicity of a given target by utilizing these precalculated molecular descriptors.  

All molecules are  in either 2D sdf format or SMILE string, and their corresponding toxicity endpoints are available on the TEST website. It should be noted that we are particularly interested in predicting quantitative toxicity endpoints so other data sets that contain qualitative endpoints or physical properties were not used. Moreover, different toxicity endpoints have different units. The units of LC$_{50}$, LC$_{50}$-DM, IGC$_{50}$ endpoints are $-\log_{10}({\mathrm{T \hspace{.02in} mol/L}})$, where {\rm T} represents corresponding endpoint. For LD$_{50}$ set, the units are $-\log_{10}({\mathrm{LD_{50} \hspace{.02in} mol/kg}})$. Although the units are not exactly the same, it should be pointed out that no additional attempt was made to rescale the values since  endpoints are of the same magnitude order.  These four data sets also differ in their sizes, ranging from hundreds to thousands, which essentially challenges the robustness of our methods. A detailed statistics table of four datasets is presented below:

\begin{table}[!ht]
\centering
\caption{Statistics of data sets }
\begin{tabular}{l|c|c|c|c|c}
\hline
  & Total \# of mols &  Train set size & Test set size & Max value & Min value\\ \hline
LC$_{50}$ set & 823  & 659  & 164  & 9.261 & 0.037\\
LC$_{50}$-DM  set & 353 & 283 & 70  & 10.064 & 0.117  \\
IGC$_{50}$ set & 1792 & 1434 & 358  & 6.36 & 0.334 \\
LD$_{50}$ set & 7413 (7397) & 5931 (5920) & 1482 (1477)  & 7.201 & 0.291 \\ \hline
\end{tabular}
\label{train_result}
\end{table} 

The number inside the parenthesis indicates the actual number of molecules that we use for developing models in this work. Note that for the first three datasets (i.e., LC$_{50}$, LC$_{50}$-DM and IGC$_{50}$ set), all molecules were properly included.  However, for LD$_{50}$ set, some molecules involved element {\rm As} were dropped out due to force field failure. Apparently, the TEST tool encounters  a similar problem  since results from two TEST models are unavailable, and the coverage (fraction of molecules predicted) from various TEST models is always smaller than one. The overall coverage of our models is always higher than  that of TEST models, which indicates a wider applicable domain of our models.

\subsection{Parameter selection}  \label{CNN_arch}

Due to the large number of adjustable parameters, it is very time consuming to optimize all possible parameter combinations. Therefore we tune parameters within a reasonable range and subsequently evaluate their performances. The parameters that we try to optimize and their range are listed below:
\begin{itemize}
\item Dropout rate $\in \{0, 0.3, 0.5\}$.
\item Number of hidden neurons in each fully-connected layer $\in \{500, 1000, 2000\}$, except for specified.
\item Number of layers in deep architecture $\in \{2, 3, 4\}$
\item SGD Momentum $m$ $\in \{0.85, 0.90, 0.95 \}$.
\item Learning rate $l$ $\in \{0.005, 0.010, 0.015 \}$.
\item $L^2$ decay rate $\beta$ $\in \{0, 10^{-6}, 10^{-4}, 10^{-2}\}$.
\end{itemize}
A more detailed discussion on parameter selection is given in Section \ref{sec:discussion}

\subsection{Evaluation criteria} 

Golbraikh {\it et al.}  \cite{golbraikh:2003} proposed a protocol to determine if a QSAR model has a predictive power. 
\begin{equation} \label{p0}
q^2 > 0.5,
\end{equation}
\begin{equation} \label{p1}
R^2 > 0.6, 
\end{equation}
\begin{equation} \label{p2}
\frac{R^2-R_0^2}{R^2} < 0.1 
\end{equation}
\begin{equation} \label{p3}
 0.85 \le k \le 1.15
\end{equation}
where $q^2$ is the  squared leave one out correlation coefficient for the training set, $R^2$ is the squared Pearson correlation coefficient between the experimental and predicted toxicities for the test set, $R_0^2$ is the squared  correlation coefficient between the experimental and predicted toxicities for the test set with the $y$-intercept being set to zero so that the regression is given by $Y=kX$. 
In addition to (\ref{p1}), (\ref{p2}) and (\ref{p3}), the prediction performance will also be evaluated in terms of root mean square error (RMSE) and mean absolute error (MAE). The prediction coverage, or fraction of chemical predicted, of corresponding methods is also taken into account since the prediction accuracy can  be increased by reducing the prediction coverage.

\section{Results} \label{sec:results}

In this section, we carry out our predictions by using four DL methods, i.e., ST-DNN, ST-CNN, MT-DNN and MT-CNN,  and two non-CNN methods, namely,  RF and GBDT. The performances of these methods are compared with those of QSAR approaches used in the development of TEST software \cite{test_guide}. For the quantitative toxicity endpoints that we are particularly interested in, a variety of methodologies were tested and evaluated  \cite{test_guide}, including hierarchical method \cite{martin:2008}, FDA method, single model method, group contribution method \cite{martin:2001} and nearest neighbor method.    
 
As for ensemble models (RF and GBDT), the training procedure follows the traditional QSAR pipeline. The raw feature vector of length $330$, without being classified into different element channels, is used as the input. A particular model is then trained to predict the corresponding toxicity endpoint. Although not all descriptors are necessary or informative, we do not perform feature selection to reduce the feature dimension as ensemble methods naturally take overfitting into account via operations such as bagging. However, the deep neural network training is slightly different. To extract abstract features from different element categories, the raw feature vector is reshaped to $11\times30$ (See \ref{sec:feature}) before fed into the network. The dropout layer and regularization of weight matrix on loss function help to deal with overfitting and feature selection. 

Note that except for specifically mentioned, all our results shown in following tables are the average outputs of 50 numerical experiments. Specifically, to eliminate randomness in neural network training, we built 50 models for each set of parameters and then used their average output as our final prediction. New consensus predictions are built by averaging  the results of RF, GBDT, MT-DNN and MT-CNN.

\subsection{Feathead minnow LC$_{50}$ test set}

The feathead minnow LC$_{50}$ set was randomly divided into a training set (80\% of the entire set) and a test set (20\% of the entire set), based on which a variety of TEST models were built. Table \ref{LC50_results} shows the performances of five TEST models, the TEST consensus obtained by the average of all independent TEST predictions,  six proposed methods and new consensus obtained from averaging over present RF, GBDT, MT-DNN and MT-CNN results. TEST consensus gives the best prediction \cite{test_guide} among TEST results, reporting a correlation coefficient of 0.728 and RMSE of 0.768 log(mol/L). When our element and molecular features are used,  RF and GBDT approaches show improvement over TEST consensus prediction,  while ST deep learning methods are outperformed. Our MT-CNN gives the best prediction over all other independent methods in terms of correlation coefficient, RMSE and MAE.   

Motivated by the good performance of TEST consensus results,  we also construct a ``New consensus'' prediction. As shown in the table, New consensus prediction has the highest correlation coefficient, and lowest RMSE and MAE among all approaches.  

\begin{table}[!ht]
\centering
\caption{Comparison of prediction results for the fathead minnow LC$_{50}$ test set.}
\begin{tabular}{c|c|c|c|c|c|c}
\hline
Method & $R^2$ & $\frac{R^2-R_0^2}{R^2}$ & $k$ & RMSE & MAE & Coverage \\ \hline
Hierarchical \cite{test_guide} & 0.710 & 0.075 & 0.966 & 0.801 & 0.574 & 0.951 \\ 
Single Model \cite{test_guide} & 0.704 & 0.134 & 0.960 & 0.803 & 0.605 & 0.945 \\ 
FDA  \cite{test_guide} & 0.626 & 0.113 & 0.985 & 0.915 & 0.656 & 0.945 \\ 
Group contribution \cite{test_guide} & 0.686 & 0.123 & 0.949 & 0.810 & 0.578 & 0.872 \\
Nearest neighbor  \cite{test_guide}& 0.667 & 0.080 & 1.001 & 0.876 & 0.649 & 0.939 \\
TEST consensus   \cite{test_guide} & 0.728 & 0.121 & 0.969 & 0.768 & 0.545 & 0.951 \\\hline
RF & 0.733 & 0.010 & 1.009 & 0.768 & 0.558  & 1.000  \\
GBDT  & 0.739 & 0.000 & 1.004 & 0.749  & 0.542 & 1.000 \\ 
ST-DNN & 0.628 & 0.021 & 1.001 & 0.910 & 0.624 & 1.000 \\
ST-CNN & 0.622 & 0.024 & 1.000 & 0.918 & 0.628 & 1.000 \\
MT-DNN & 0.738 & 0.005 & 1.017 & 0.754 & 0.524 & 1.000 \\
MT-CNN & 0.748 & 0.002 & 1.011 & 0.738 & 0.508 & 1.000 \\ 
New consesus & 0.776 & 0.005 & 1.015 & 0.699 & 0.492 & 1.000 \\ \hline
\end{tabular}
\label{LC50_results}
\end{table}

\begin{figure}[!ht]
\small
\centering
\includegraphics[scale=0.6]{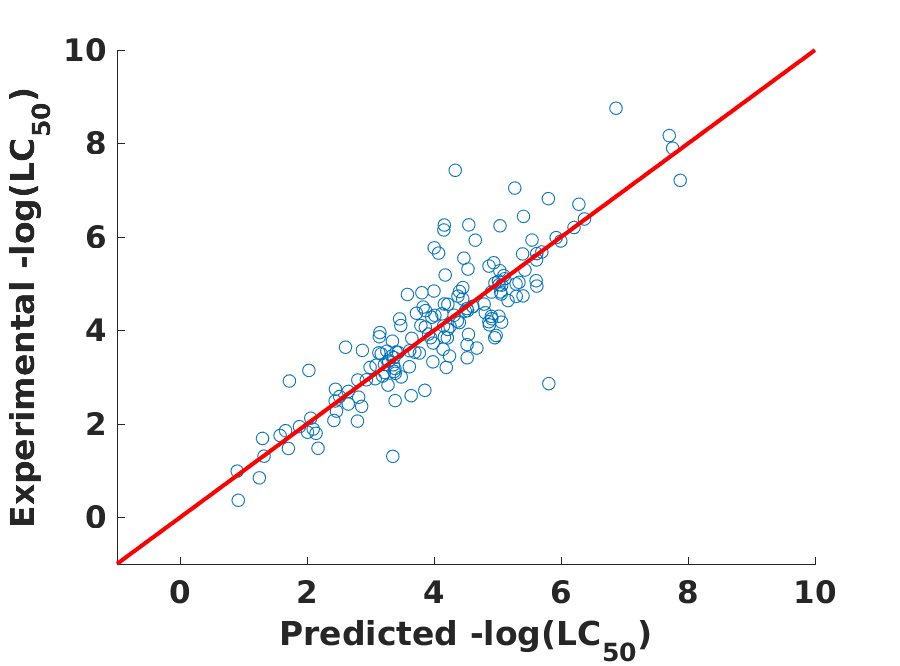}
\caption{Experimental verseus predicted values by MT-CNN  for the fathead minnow LC$_{50}$ set. }
\label{lc50_plot}
\end{figure}

Figure \ref{lc50_plot} shows the values predicted by MT-CNN  for the feathead minnow LC$_{50}$ test set against experimental values. Our predictions are  quite bias free. 

\subsection{Daphnia magna LC$_{50}$ test set} 

The Daphinia Magna LC50 set is the smallest in terms of set size, with 283 training molecules and 70 test molecules, respectively. However, it brings difficulties to building robust QSAR models given the relatively large number of descriptors. Indeed, five independent models in TEST software give significantly different predictions, as indicated by RMSEs shown in Table \ref{LC50DM_results} ranging from 0.810 to 1.190 log units. Though the RMSE of Group contribution is the smallest, its coverage is only  0.657 \% which largely restricts this method's applicability. Additionally, its $R^2$ value is inconsistent with its RMSE and MAE.  Since Ref. \cite{test_guide} states that ``The consensus method achieved the best results in terms of both prediction accuracy and coverage'', these usually low RMSE and MAE values might be typos.

We also notice that our non-multitask models result in very large deviation from experimental values.   Nevertheless, it is encouraging to see that   MT-CNN  outperforms other independent methods and TEST consensus with an RMSE of 0.872 log units and 100 \% coverage. This result suggests that by learning related problems jointly and extracting shared information from different data sets, MT-CNN architecture can simultaneously perform multiple prediction tasks and make improvements over singletask models. Our new consensus prediction constructed from RF, GBDT, MT-DNN and MT-CNN offers slightly worse accuracy than that our  MT-CNN, due to the poor performance of RF and GBDT. 

\begin{figure}[!ht]
\small
\centering
\includegraphics[scale=0.6]{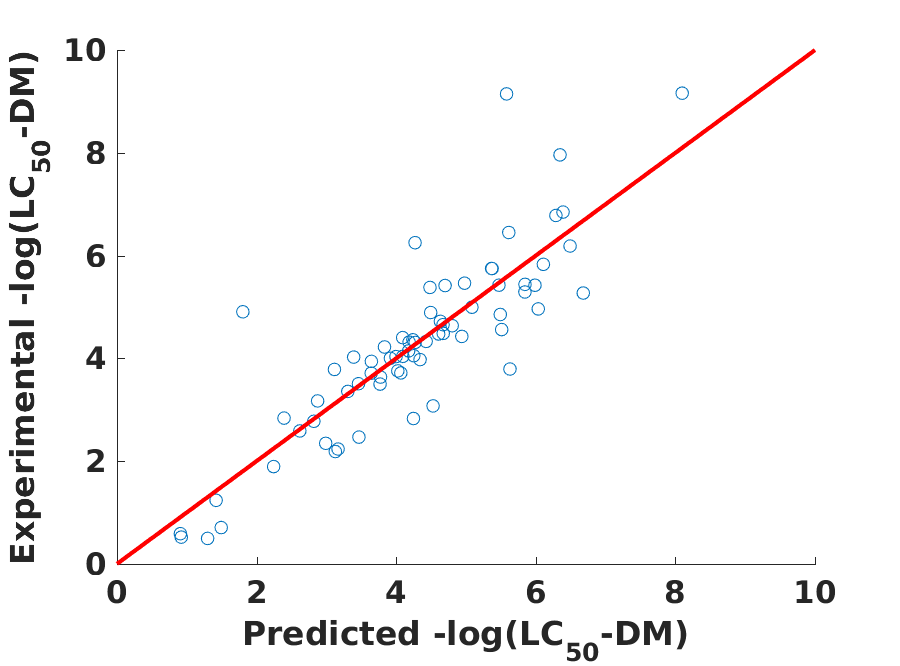}
\caption{Experimental verseus predicted values by MT-CNN  for the Daphnia magna LC$_{50}$ set. }
\label{fig:lc50dm_plot}
\end{figure}

A scatter plot of MT-CNN predictions  against experimental values can be found in Fig. \ref{fig:lc50dm_plot}. Clear, large deviations from two erroneous predictions contribute  a relatively large RMSE compared with that of other test sets.     

\begin{table}[!ht]
\centering
\caption{Comparison of prediction results for the Daphnia magna LC$_{50}$ test set.}
\begin{tabular}{c|c|c|c|c|c|c}
\hline
Method   & $R^2$ & $\frac{R^2-R_0^2}{R^2}$ & $k$ & RMSE & MAE & Coverage \\ \hline
Hierarchical \cite{test_guide} & 0.695 & 0.151 & 0.981 & 0.979 & 0.757 & 0.886 \\ 
Single Model  \cite{test_guide}& 0.697 & 0.152 & 1.002 & 0.993 & 0.772 & 0.871 \\ 
FDA  \cite{test_guide}& 0.565 & 0.257 & 0.987 & 1.190 & 0.909 & 0.900 \\ 
Group contribution  \cite{test_guide}& 0.671 
& 0.049 & 0.999 & 0.803$^a$ & 0.620$^a$ & 0.657 \\
Nearest neighbor \cite{test_guide} & 0.733 & 0.014 & 1.015 & 0.975 & 0.745 & 0.871 \\
TEST consensus   \cite{test_guide}& 0.739 & 0.118 & 1.001 & 0.911 & 0.727 & 0.900 \\\hline
RF  & 0.451 & 0.022 & 0.962  & 1.287 & 0.967 & 1.000  \\
GBDT  & 0.504 & 0.025 & 0.971 & 1.228 & 0.919 & 1.000 \\ 
ST-DNN & 0.444 & 0.658 & 0.897 & 1.584 & 1.034 & 1.000 \\
ST-CNN & 0.442 & 0.720 & 0.890 & 1.611 & 1.036 & 1.000\\
MT-DNN & 0.727 & 0.001 & 1.003 & 0.900 & 0.601 & 1.000 \\
MT-CNN & 0.743 & 0.001 & 1.002 &0.872 & 0.576 & 1.000 \\ 
New consensus & 0.684 & 0.009 & 0.995 & 0.977 & 0.729 & 1.000 \\ \hline
\end{tabular}\\
{$^a$ these values are inconsistent with $R^2=0.671$.}
\label{LC50DM_results}
\end{table}

\subsection{Tetraphymena pyriformis IGC$_{50}$ test set} 

\begin{figure}[!ht]
\centering
\includegraphics[scale=0.6]{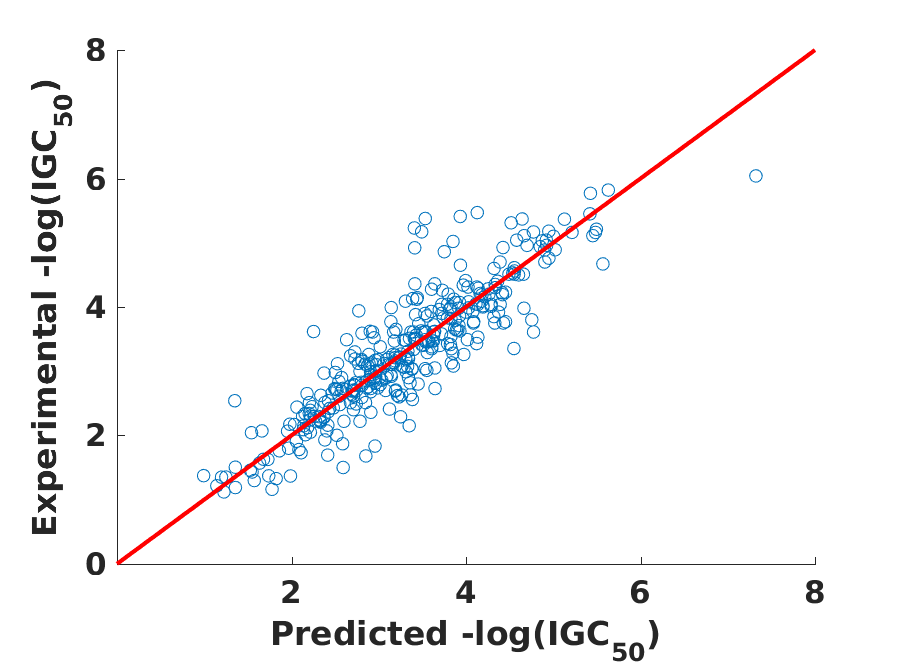}
\caption{Experimental verseus predicted values by MT-CNN  for the Tetraphymena Pyriformis IGC$_{50}$ test set. }
\label{fig:igc50_plot}
\end{figure}

IGC$_{50}$ set is the second largest QSAR toxicity set that we want to predict. The diversity of molecules of in IGC$_{50}$ set is low and the coverage of TEST  methods is relatively high compared to   previous LC$_{50}$ sets. As shown in Table \ref{IGC50_results}, 
the Single Model results are not listed.  The Test consensus prediction again yields the best result for TEST software. As for our models, the accuracies span a range of 0.112 (0.452 to 0.564) log units. Compared to all independent methods, the MT-CNN model achieves the best prediction, with an RMSE of 0.452 log units. The low variation of RMSEs of our models indicate that our element and molecular  descriptors are insensitive to machine learning algorithms and therefore able to model this specific toxicity endpoint (IGC$_{50}$) in a robust way. It is noted that 
our new consensus prediction constructed from RF, GBDT, MT-DNN and MT-CNN delivers the best result on all measures. 
The experimental values versus MT-CNN predicted values plot is given in Fig. \ref{fig:igc50_plot}.   

\begin{table}[!ht]
\centering
\caption{Comparison of  prediction results for the Tetraphymena Pyriformis IGC$_{50}$ test set.}
\begin{tabular}{c|c|c|c|c|c|c}
\hline
Method & $R^2$ & $\frac{R^2-R_0^2}{R^2}$ & $k$ & RMSE & MAE & Coverage \\ \hline
Hierarchical  \cite{test_guide}& 0.719 & 0.023 & 0.978 & 0.539 & 0.358 & 0.933 \\ 
FDA  \cite{test_guide}& 0.747 & 0.056 & 0.988 & 0.489 & 0.337 & 0.978 \\ 
Group contribution  \cite{test_guide}& 0.682 & 0.065 & 0.994 & 0.575 & 0.411 & 0.955 \\
Nearest neighbor  \cite{test_guide}& 0.600 & 0.170 & 0.976 & 0.638 & 0.451 & 0.986 \\
TEST consensus  \cite{test_guide} & 0.764 & 0.065 & 0.983 & 0.475 & 0.332 & 0.983 \\\hline
RF  & 0.702 & 0.005 & 0.990 & 0.539 & 0.382 & 1.000  \\
GBDT  & 0.749 & 0.006 & 0.989 & 0.497 & 0.343 & 1.000 \\ 
ST-DNN & 0.694 & 0.034 & 0.980 & 0.564 & 0.349 & 1.000 \\
ST-CNN & 0.689 & 0.039 & 0.979 & 0.571 & 0.352 & 1.000 \\
MT-DNN & 0.768 & 0.012 & 0.993 & 0.483 & 0.329 & 1.000 \\
MT-CNN & 0.791 & 0.003 & 0.997  & 0.452 & 0.314 & 1.000 \\ 
New consensus & 0.793 & 0.000 & 0.996 & 0.447 & 0.310 & 1.000 \\ \hline
\end{tabular}
\label{IGC50_results}
\end{table}

\subsection{Oral rat LD$_{50}$ test set} 

The oral rat LD$_50$ set contains the largest molecule pool with 7413 compounds. However, none of methods is able to provide a 100\% coverage of this data set. The results of single model method or group contribution method were not properly built for the entire set \cite{test_guide}.  It was noted that  LD$_{50}$ values of this data set are relatively difficult to predict as they have a higher experimental uncertainty \cite{zhuqsar:2009}. As shown in Table \ref{LD50_results}, results of two TEST approaches, i.e., Single Model and Group contribution, were not reported for this problem. The TEST consensus result improves overall prediction accuracy of other TEST methods by about 10 \%. 

\begin{figure}[!ht]
\small
\centering
\includegraphics[scale=0.6]{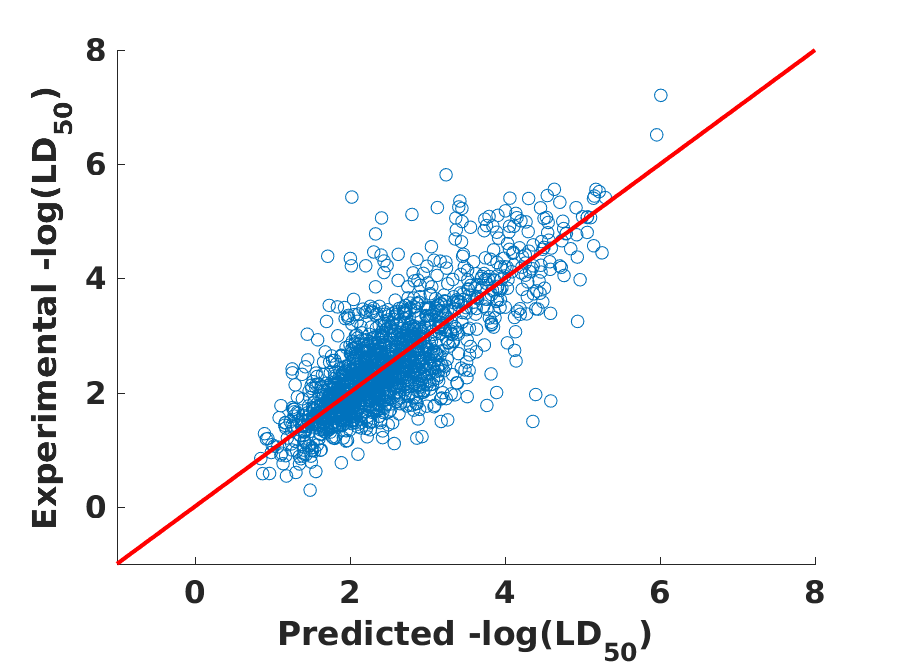}
\caption{Experimental verseus predicted values by MT-CNN for the oral rat LD$_{50}$ test set. }
\label{fig:ld50_plot}
\end{figure}

Meanwhile, our predictions are also relatively stable for this particular set. The RMSEs of non multitask models fall within the interval of [0.620, 0.673] log units  and do not essentially fluctuate. As expected, the MT-CNN model gives a higher accuracy (RMSE of 0.604 log units) than our other independent models while performing slightly worse than TEST consensus prediction (RMSE of 0.594 log units). 
However, our new consensus prediction obtained from RF, GBDT, MT-DNN and MT-CNN results outperforms all TEST methods and our own approaches with an RMSE of 0.582 log units.   
The  scatter plot of MT-CNN prediction is given in  Fig. \ref{fig:ld50_plot}. Clear, MT-CNN predictions are very balanced. 

\begin{table}[!ht]
\centering
\caption{Comparison of  prediction results for the Oral rat LD$_{50}$ test set.}
\begin{tabular}{c|c|c|c|c|c|c}
\hline
Method & $R^2$ & $\frac{R^2-R_0^2}{R^2}$ & $k$ & RMSE & MAE & Coverage \\ \hline
Hierarchical  \cite{test_guide}& 0.578 & 0.184 & 0.969 & 0.650 & 0.460 & 0.876 \\ 
FDA  \cite{test_guide}& 0.557 & 0.238 & 0.953 & 0.657 & 0.474 & 0.984 \\ 
Nearest neighbor \cite{test_guide} & 0.557 & 0.243 & 0.961 & 0.656 & 0.477 & 0.993 \\
TEST consensus  \cite{test_guide} & 0.626 & 0.235 & 0.959 & 0.594 & 0.431 & 0.984 \\\hline
RF & 0.592 & 0.010 & 1.000 & 0.620 & 0.464 & 0.997  \\
GBDT  & 0.578 & 0.002 & 0.991  & 0.627 & 0.476 & 0.997 \\ 
ST-DNN & 0.550 & 0.039 & 0.987 & 0.662 & 0.456 & 0.997 \\
ST-CNN & 0.540 & 0.050 & 0.986 & 0.673 & 0.456 & 0.997 \\
MT-DNN & 0.586 & 0.023 & 0.994 & 0.630 & 0.453 & 0.997 \\
MT-CNN & 0.612 & 0.008 & 0.998 & 0.604 & 0.437 & 0.997 \\ 
New consensus & 0.636 & 0.001 & 1.002 & 0.582 & 0.432 & 0.997 \\ \hline
\end{tabular}
\label{LD50_results}
\end{table}

\section{Discussion} \label{sec:discussion}

In this section, we present detailed discussion about optimized MT-CNN versus other methods. The adjustable parameters range from number of hidden layers to $L^2$ decay rates, and they can be found in Section \ref{CNN_arch}.  The final proposed parameters for the trained network are listed in Table \ref{parameters} and all results shown after are based on this default parameter set.
\begin{table}[!ht]
\centering
\caption{Proposed parameters for MT-CNN.}
\begin{tabular}{c|c}
\hline
Number of hidden layers & 3 \\
Number of neurons on each layer & 2000, 1000, 1000 \\ 
Dropout rate & 0 \\
$L^2$ decay rate & $10^{-6}$ \\
Learning rate & 0.010 \\
SGD momentum & 0.90 \\
\hline 
\end{tabular}
\label{parameters}
\end{table}

\subsection{Convolutional deep learning vs non-convolutional deep learning} 

\begin{figure}[!ht]
\centering
\includegraphics[scale=0.6]{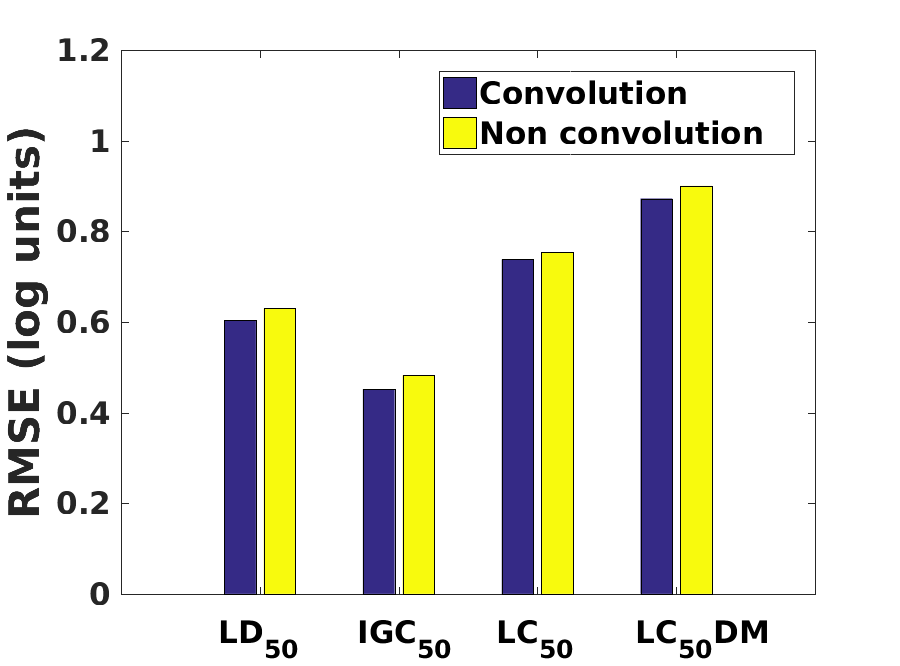}
\caption{The results of deep learning architectures with or without convolutional operations. }
\label{fig:conv_plot}
\end{figure}

An essential idea of our feature construction process is to build element-specific `channels' to extract information by element types. Our convolution CNN architecture takes advantage of  convolution operations to capture higher level representation of toxicity endpoints. As expected, for essentially all models that we have trained, those that use matrix features as input give better performances, provided that other parameters remain the same. Figure \ref{fig:conv_plot} depicts a comparison  between MT-CNN and non-convolutional MT-DNN.  Both architectures are optimized. One can clearly see that MT-CNN architecture outperforms the non-convolutional MT-DNN model.

We have also tried to divide our features into three types, i.e., charge, area and energy, and then, restrict the convolution within each type of features. The resulting MT-CNN architecture behaves similarly to the present one.

\subsection{Multi-task CNN vs single-task CNN} 

One major motivation for multi-task learning is to gain information from related tasks. Given a set of parameters, multi-task learning is capable of predicting results for 4 different QSAR tasks. It turns out that the size of a training set plays an important role on the extent to which our predictions can benefit from multi-task learning. For medium to large sized sets, namely LC$_{50}$, IGC$_{50}$ and LD$_{50}$ sets,  multi-task architecture offers a substantial improvement over single-task models. Nevertheless, when a limited number of molecules is available (LC$_{50}$-DM set), the results significantly favor multi-task models, which indicates that there exists shared information across related tasks that can benefit prediction accuracy. A detailed comparison of multi-task deep learning models and single-task deep learning models can be found in Section \ref{sec:results}.

\subsection{The effect of parameter selection for  MT-CNN.} 

\begin{figure}[ht!]
\centering
\begin{subfigure}{0.5\textwidth}
\centering
\includegraphics[width=0.95\textwidth]{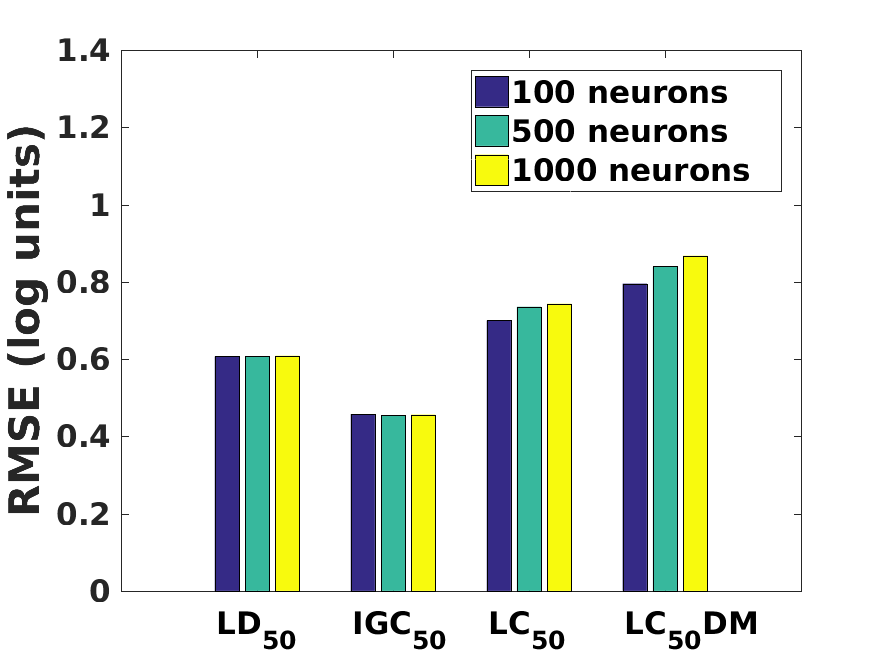}
\caption{Results for different number of neurons on each layer}
\end{subfigure}%
\begin{subfigure}{0.5\textwidth}
\centering
\includegraphics[width=0.95\textwidth]{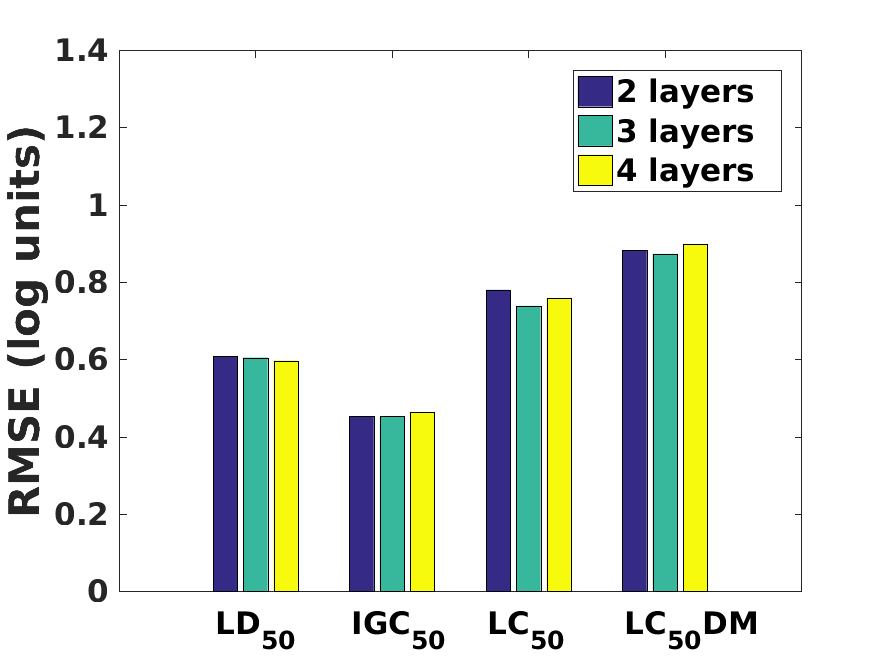}
\caption{Results for different number of layers}
\end{subfigure}%
\caption{A direct comparison of MT-CNN architectures with different number of neurons per layer (a) and different number of layers (b).}
\label{fig:cmp1}
\end{figure}

Different sets of parameters were explored to optimize the proposed MT-CNN architecture.  The first parameter that we would like to tune is the number of neurons per layer. We construct neural networks with 100, 500, and 1000 neurons on each of the three layers (here the number of layers is set to 3 for simplicity). It is worthy to mention that the ensemble result of the smallest number of neurons, namely 100 neurons per layer,  gives the best results as Fig. \ref{fig:cmp1} shows. The underlying reason is that each individual model generates predictions with higher bias and variance, which are offset by averaging. In contrast, since the individual model of larger networks is of lower bias and variance, the average turns out to be slightly worse. Next we built networks with 2, 3 and 4 hidden layers followed by fully-connected output layers. As shown in Fig. \ref{fig:cmp1}, increasing the number of hidden layers does not necessarily improve the prediction accuracy, indicating that sufficient neurons on each layer have successfully captured the underlying representation of toxicity endpoints.

\begin{figure}[ht!]
\centering
\begin{subfigure}{0.5\textwidth}
\centering
\includegraphics[width=0.95\textwidth]{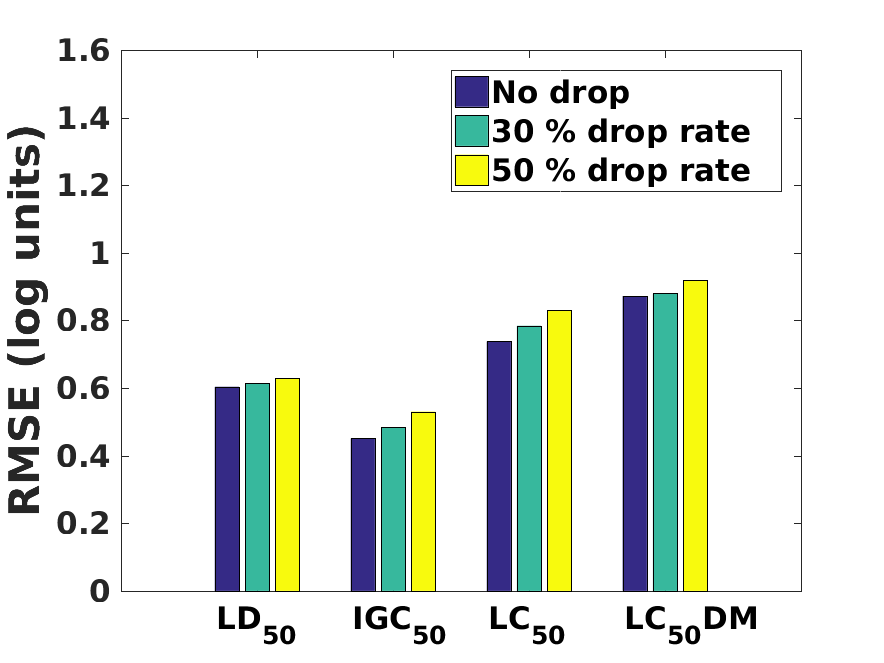}
\caption{Results for different drop rates}
\end{subfigure}%
\begin{subfigure}{0.5\textwidth}
\centering
\includegraphics[width=0.95\textwidth]{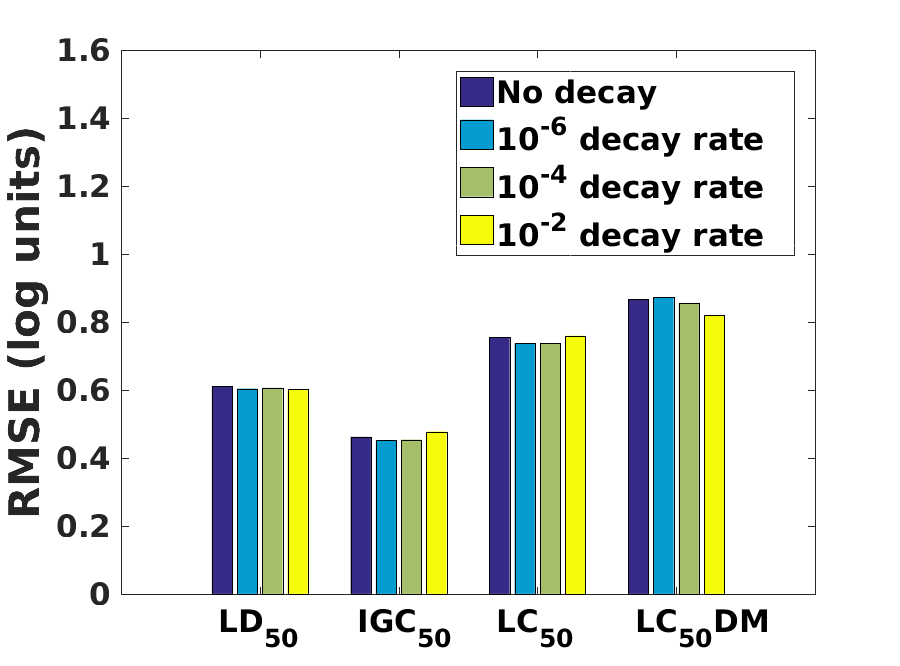}
\caption{Results for different $L^2$ decay rates}
\end{subfigure}%
\caption{A direct comparison of MT-CNN architectures with different drop rates and $L^2$ decay rates.}
\end{figure}

Secondly, we would like to explore the impact of dropout rate and $L^2$ decay on MT-CNN predictions, which are two ways to avoid network overfitting. When no dropout or low dropout rate is applied, the results do not have significant differences. When dropout rate is increased to 50 \%, however, the model gives worse predictions as compared to other models, which indicates a high information loss. As for $L^2$ decay, it is somehow surprising to see that there does not seem to exist noticeable improvement across different parametrizations. A possible explanation is that learning from multiple data sets helps to adjust weights of networks within a reasonable range and prevent weights from growing out of control. 

\begin{figure}[ht!]
\centering
\begin{subfigure}{0.5\textwidth}
\centering
\includegraphics[width=0.95\textwidth]{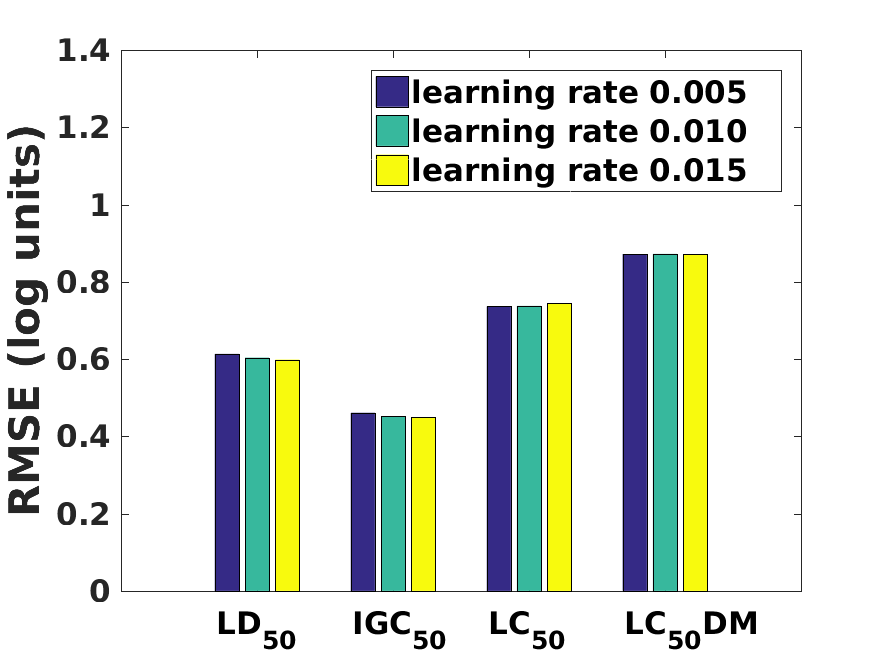}
\caption{Results for different learning rates}
\end{subfigure}%
\begin{subfigure}{0.5\textwidth}
\centering
\includegraphics[width=0.95\textwidth]{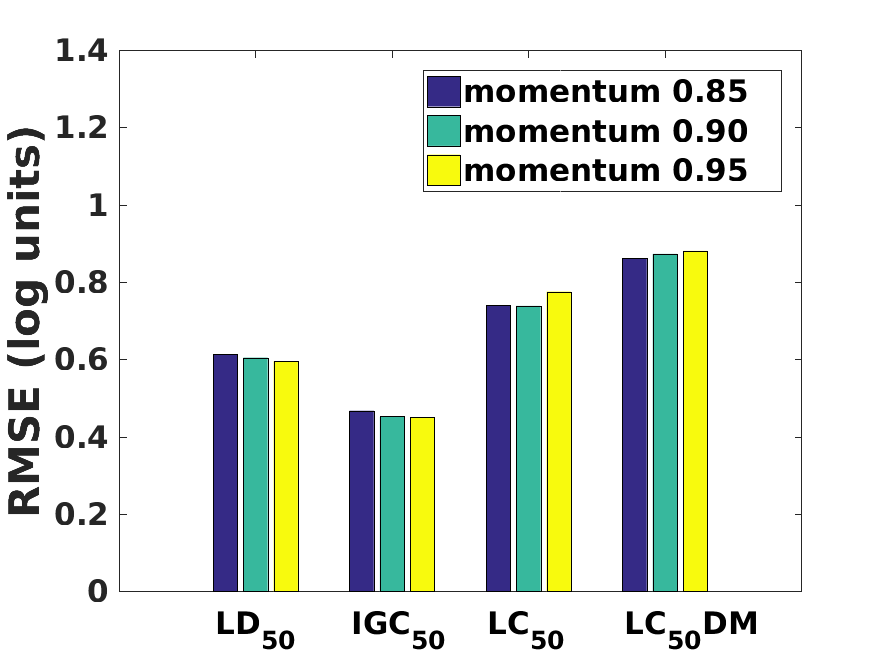}
\caption{Results for different SGD momentum}
\end{subfigure}%
\caption{A direct comparison of architectures with different learning rates and SGD momentum.}
\label{fig:cmp4}
\end{figure}

Third, it is also interesting to know how learning rate and SGD momentum can affect the overall performance of models. As Fig. \ref{fig:cmp4} shows, there is no significant difference between different combinations of learning rates and SGD momentum. It indicates that after a sufficient number of training epochs, the gradient of the network parameters starts to converge to the same location and, as a result, the trained networks yield very similar predictions.

\section{Conclusion} \label{sec:conclusion}

Toxicity refers to the degree of damage a substance on an organism,  such as an animal, bacterium, or plant, and can be measured experimentally on various targets. However, experimental measurement of toxicity is typically expensive and time consuming, in addition to potential ethic concerns. Theoretical prediction of toxicity has become a useful alternative in  pharmacology and environmental science. A wide variety of methods has been developed for toxicity prediction in the past. The performances of these methods depend  not only on the descriptors, but also on machine learning algorithms, which makes the model evaluation a difficult task. In particular, typical deep learning algorithms appear to offer little or no conceptual  understanding of their prediction results. In this work, we construct a common set of  microscopic features based on well established physical models to examine the performance of various deep neural network (DNN) and convolutional neural network (CNN) approaches (i.e., multi-task (MT) and single-task (ST) DNN and CNN models) and two ensemble  methods (i.e., random forest (RF) and gradient boosting decision tree (GBDT)). Comparison has also been made to the state-of-art approaches given in the literature \href{https://www.epa.gov/chemical-research/toxicity-estimation-software-tool-test }{Toxicity Estimation Software Tool} (TEST) \cite{test_guide}  listed by United States Environmental Protection Agency. 
   
Four toxicity data sets, i.e., 96 hour fathead minnow LC$_{50}$ data set (LC$_{50}$ set), 48 hour Daphnia magna LC$_{50}$ data set (LC$_{50}$-DM set), 40 hour Tetrahymena pyriformis IGC$_{50}$ data set (IGC$_{50}$ set), and oral rat LD$_{50}$ data set (LD$_{50}$ set), are used in the present TEST study.  Our numerical experiments indicate  the follows. First, ST neural networks, including ST-CNN and ST-DNN, have similar performance over all the data sets. Similarly two  ensemble methods, i.e., RF and GBDT, deliver a similar level of accuracy. Two MT models, MT-DNN and MT-CNN also have a similar level of accuracy, while MT-CNN slightly outperforms MT-DNN in all cases. Second, RF and GBDT are typically more accurate than ST-CNN and ST-DNN. However, MT deep learning algorithms are typically  more accurate than ensemble methods. Third, MT-CNN outperforms all the other independent methods in terms of accuracy for all test sets studied.   Finally,  new state-of-the-art results for these toxicity problems are established either by MT-CNN or by   new consensus  predictions built from RF, GBDT, MT-DNN and MT-CNN results. 

It is worthy to note that since the proposed methods are built only from three types of  microscopic features (i.e., charges, surface areas and free energies), they  have the highest coverage among all the existing methods compared, indicating their boarder applicability to practical toxicity analysis and prediction.   

\section*{Acknowledgments}

This work was supported in part by NSF Grant  IIS-1302285  and MSU Center for Mathematical Molecular Biosciences Initiative.

\bibliographystyle{unsrt}
\bibliography{refs}

\end{document}